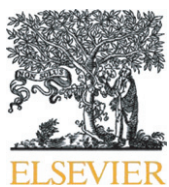
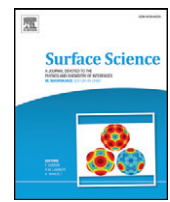
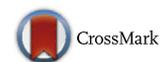

# From atoms to steps: The microscopic origins of crystal evolution

Paul N. Patrone [a,b,*], T.L. Einstein [c], Dionisios Margetis [d]

[a] Center for Nanoscale Science and Technology, National Institute of Standards and Technology, Gaithersburg, MD 20899-1070 USA
[b] Institute for Mathematics and its Applications, University of Minnesota, Minneapolis, MN 55455 USA
[c] Department of Physics and Condensed Matter Theory Center, University of Maryland, College Park, MD 20742-4111 USA
[d] Department of Mathematics, and Institute for Physical Science and Technology, and Center for Scientific Computation and Mathematical Modeling, University of Maryland, College Park, MD 20742, USA

## ABSTRACT

The Burton–Cabrera–Frank (BCF) theory of crystal growth has been successful in describing a wide range of phenomena in surface physics. Typical crystal surfaces are slightly misoriented with respect to a facet plane; thus, the BCF theory views such systems as composed of staircase-like structures of steps separating terraces. Adsorbed atoms (adatoms), which are represented by a continuous density, diffuse on terraces, and steps move by absorbing or emitting these adatoms. Here we shed light on the microscopic origins of the BCF theory by deriving a simple, one-dimensional (1D) version of the theory from an atomistic, kinetic restricted solid-on-solid (KRSOS) model without external material deposition. We define the *time-dependent* adatom density and step position as appropriate ensemble averages in the KRSOS model, thereby exposing the non-equilibrium statistical mechanics origins of the BCF theory. Our analysis reveals that the BCF theory is valid in a *low adatom-density* regime, much in the same way that an ideal gas approximation applies to dilute gasses. We find conditions under which the surface remains in a low-density regime and discuss the microscopic origin of corrections to the BCF model.

Published by Elsevier B.V.

## 1. Introduction

The controlled growth and etching of crystals is an important process that has applications in a variety of settings, including the fabrication of microprocessors, quantum dots, and nanowires, to name a few [1]. Since these processes often involve the assembly of structures at the nanoscale, where the misplacement of a few atoms can have large effects, it is important to develop theoretical models that improve our understanding and control of such evolution processes. In particular, mesoscale models at the 10 nm–100 nm range have gained considerable attention due to the fact that they provide a computationally tractable means to study discrete elements (e.g. defects) of nanoscale systems without tracking every individual atom. Formulating methods to connect atomistic and mesoscale models of crystalline surfaces is a critical task in theoretical physics [2].

In this paper, we discuss the derivation of one such mesoscale model, the Burton–Cabrera–Frank (BCF) theory [3], in a one-dimensional (1D) setting. In 1951, BCF postulated an important mechanism of surface evolution that came to be known as *step flow*. They viewed crystal surfaces as composed of staircase-like structures, i.e. systems of steps separating terraces (cf. Fig. 1). Adsorbed atoms (adatoms) diffuse on the terraces until arriving at and then attaching to a step. Such attachment (and the corresponding detachment) processes cause the steps to move, which can lead to large-scale morphological changes in the crystal over long enough times.

Mathematically, BCF chose to formulate this microscopic picture in terms of a Stefan-type free boundary problem [4]. Adatoms are represented by a density *c* that obeys a diffusion equation; boundary conditions at the step account for the physics of the attachment/detachment processes, and the velocity of a step, which is a free (or movable) boundary, is proportional to the net flux of adatoms arriving at the step [3]. While this perspective is physically appealing, BCF did not derive their model from an atomistic theory of surface diffusion. Thus, many questions remain about the underlying assumptions and limitations of the BCF theory. Is there an unambiguous relationship between the mesocale parameters of the BCF theory and the atomistic parameters describing adatom motion? How far from equilibrium can a system be and still be well described by the BCF theory? How does the theory break down?

Our goal in this paper is to heuristically answer these questions by deriving a one-step, 1D version of the BCF theory from an idealized, atomistic model of surface diffusion. Our discussion here is largely formal; a full treatment of this problem requires a significant foray into set theory and functional analysis.[1] Nonetheless, a key observation allows for a simplification that makes the problem easily accessible: experiments and simulations show that the number of adatoms on many surfaces is typically very small.[2] Thus, we choose as our starting point a *one-adatom* kinetic restricted solid-on-solid (KRSOS) model. In the more general context of a kinetic solid-on-solid model [7], the surface configuration is represented as an array of columns of integer heights by exclusion of overhangs; and the governing master equation describes the probability of finding each configuration according to certain rules for the hopping of surface atoms from one site to another. Here, we pose a simplified, 1D version of this description with one adatom and one step; and invoke a master equation that describes the probability of finding the adatom at a particular location on the surface. Our simplifying assumptions preclude the possibility of step permeability (transitions in which adatoms can go from one terrace to the next without stopping at the step) [8].

---

* Corresponding author.
E-mail addresses: ppatrone@umn.edu (P.N. Patrone), einstein@umd.edu (T.L. Einstein), dio@math.umd.edu (D. Margetis).

[1] For a discussion that addresses some mathematical aspects of the full problem, see Ref. [5].

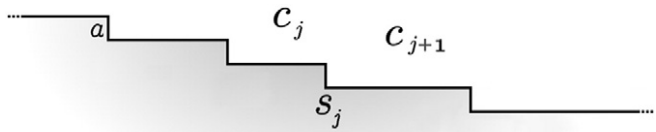

**Fig. 1.** A 1D step system with several steps (whose positions are labeled $s_j$) separating terraces; $a$ is the atomic step height. Adatoms, represented by the densities $c_j$, diffuse on each terrace. The velocity of a step is proportional to the net current of adatoms arriving at the step.

Our main task amounts to reconciling the stochastic and discrete nature of the KRSOS model with the notions of a deterministic and continuous adatom density and step position of the BCF theory. This reconciliation can be achieved by defining the adatom density and step position as *time-dependent* expectation values taken over the master equation solution; the BCF theory, plus corrections, then describes the time evolution of these averages. This approach in particular has the benefit of showing how the BCF theory extends ideas of equilibrium statistical mechanics to non-equilibrium systems.

We caution, however, that the BCF theory (as we derive it) is not always valid out of equilibrium. Indeed, an important aspect of our analysis is to determine the "near-equilibrium" conditions under which the BCF theory produces predictions consistent with the KRSOS picture. To this end, we derive a maximum principle (often found in the analysis of the heat equation [9]) to show when corrections to the BCF theory can be neglected. We also describe, but do not derive, corrections, due to *adatom correlations*, that arise from a multi-particle KRSOS model (as in Ref. [5]) and discuss the conditions under which they can be neglected; see also Refs. [6,10].

Several works have addressed questions related to the derivation of the BCF model [11–15]. Here it suffices to note that these works only derived isolated parts of the BCF theory, whereas we aim to derive all of its elements together. However, the 1D, one-step KRSOS model that we invoke brings its own set of limitations. In particular, real crystal surfaces are two-dimensional (2D), and steps usually have profiles that are not perfectly straight. These two facts greatly complicate the formulation of an atomistic model, and to the best of our knowledge, no derivation of a fully 2D BCF theory has been achieved. Nonetheless, we believe that our analysis sheds light on the key atomistic processes that give rise to the BCF theory.

The rest of the paper is organized as follows. In Section 2, we present the mathematical elements of the BCF theory. In Section 3, we present a one-particle version of our one-step, 1D KRSOS model. In Section 4, we formulate an averaging procedure by which the KRSOS master equation can be transformed into discrete, BCF-type equations, plus corrections.

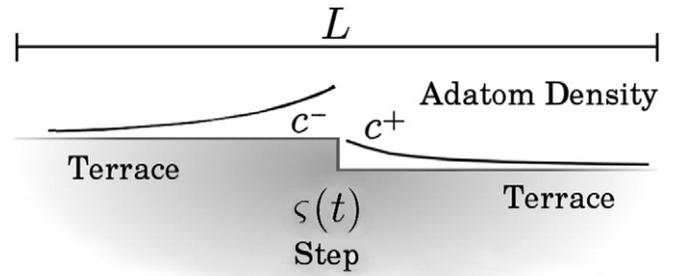

**Fig. 2.** The 1D single-step system that we consider; the step position is denoted $\varsigma(t)$. The values $c^\pm$ are the adatom densities on the right ($+$) and left ($-$) sides of the step; $L$ is the length of the system.

In Section 5, we provide a maximum principle that yields conditions under which we can neglect the corrections, and in Section 6 we take the continuum limit of the discrete BCF equations. In Section 7, we discuss corrections that arise from a multi-particle KRSOS model, and in Section 8 we present our main conclusions. Appendix A provides a proof of the maximum principle that we invoke in Section 5.

## 2. A one-step BCF model

We consider the one-step system illustrated in Fig. 2. A step at position $\varsigma(t)$ separates an upper and a lower terrace. Adatoms, represented by the density $c(x,t)$, obey the diffusion equation

$$\partial_t c(x,t) = \mathcal{D}\partial_x^2 c(x,t) \qquad \text{or} \qquad 0 \leq x < \varsigma(t), \ \varsigma(t) < x \leq L, \qquad (1)$$

where $\mathcal{D}$ is a constant diffusivity, $L$ is the length of the system, and $\partial_x \equiv \partial/\partial x$. We apply periodic boundary conditions at $x = 0$ and $x = L$.

Boundary conditions at the steps are of the form

$$\mathcal{J}_\pm = -\mathcal{D}\partial_x c\big|_\pm = \mp\kappa_\pm\left(c^\pm - c^{eq}\right), \qquad (2)$$

where $\mathcal{J}_\pm$ is the adatom flux at the right ($+$) or left ($-$) edge of the step, $\kappa_\pm$ is an attachment/detachment rate at the right ($+$) or left ($-$) edge of the step,[3] and $c^\pm$ is the adatom concentration to the right ($+$) or left ($-$) of the step; $c^{eq}$ is an equilibrium adatom concentration. Eq. (2) states that the step will emit or absorb adatoms until the densities $c^\pm$ attain their equilibrium values. The density $c^{eq}$ is generally assumed to be of the form

$$c^{eq} \propto e^{-\mu/k_B T}, \qquad (3)$$

where $\mu$ is a *step chemical potential*, i.e. the energy added to the system when an atom attaches to the step [16].

Because the step can move, we require an additional equation to describe its motion. We set the step velocity $\dot\varsigma(t)$ equal to the net current,

$$\dot\varsigma(t) = a(\mathcal{J}_- - \mathcal{J}_+), \qquad (4)$$

where $a$ is the lattice spacing. Eq. (4) is a mass conservation statement; adatoms attaching to (detaching from) a step cause it to advance (retreat).

Our goal in the remainder of this paper is to derive Eqs. (1)–(4). In particular, an important part of our analysis is to identify $\kappa_\pm$ and $\mu$ in terms of the processes in our KRSOS model.

---

[2] For an illustration of the modifications needed in the case of high adatom density, see Ref. [6].

[3] The original BCF formulation (Ref. [3]) amounts to $\kappa_\pm \to \infty$, so that $c = c^{eq}$ at the step edge.

## 3. Kinetic restricted solid-on-solid (KRSOS) model

In the context of surfaces, our KRSOS model is a special case of a stochastic lattice-gas model, which is a probabilistic representation of the system accounting for the random motion of individual atoms. Solutions to our KRSOS model are the time-dependent probabilities of finding the system in each of its atomistic configurations. Given some initial state, the model describes how the system transitions between its accessible configurations.

In general, a stochastic lattice-gas model allows for an arbitrary number of particles to move; see, for example, Refs. [5,7,12,16,17]. However, in many experiments and simulations, one finds that few lattice sites are occupied by adatoms, which instead spend most of their time attached to a step. This observation motivates a key simplification of our KRSOS model: we only consider a system in which one adatom is ever allowed to move. While this simplification may seem drastic, Ref. [5] shows that, to good approximation, the behavior of a multi-particle model is often well described by the one-particle (1-p) model that we consider here. As we show in Section 5, this 1-p KRSOS model also contains the essential physics of the BCF theory.

Thus, consider the system illustrated in Fig. 3. A surface with a single step is divided into $N$ lattice sites indexed by $j$, where $0 \leq j \leq N - 1$. For definiteness, we pick a lattice site $s_0$ and use it to define a *microscopic step position* as follows[4]: we call every site $j \neq s_0$ a "terrace site." Whenever the adatom is at any site $j \neq s_0$, the *microscopic* step is at position $s_0 - 1$; when the adatom is at site $s_0$, it becomes part of the step, whose position is then also $s_0$.[5] Our KRSOS model consists of the following set of rules that describe how the atom hops between lattice sites.

Rule 1. *The adatom can only hop to one of its two adjacent lattice sites.*
Rule 2. *The adatom hops from a terrace site to any adjacent terrace site with a probability proportional to a constant rate $D$ (described below).*
Rule 3. *The adatom hops to the step (i.e. $j = s_0$) from the left (−) or right (+) with probability proportional to an attachment rate $D\phi_\pm$ (defined below).*
Rule 4. *The adatom detaches from the step to the left (−) or right (+) with probability proportional to a detachment rate $Dk\phi_\pm$ (defined below).*

Analytically, these rules are expressed via a *master equation*, a system of ordinary differential equations describing the probabilities of finding the atom at each lattice site. If $p_j(t)$ is the probability that the atom is at site $j$, the corresponding master equation is

$$\dot{p}_j = D\left[p_{j+1} - 2p_j + p_{j-1}\right], \quad j \neq 0, s_0, s_0 \pm 1, N-1, \tag{5}$$

$$\dot{p}_{s_0 \pm 1} = D\left[k\phi_\pm p_{s_0} - (1 + \phi_\pm)p_{s_0 \pm 1} + p_{s_0 \pm 2}\right], \tag{6}$$

$$\dot{p}_{s_0} = D\left[\phi_- p_{s_0-1} - k(\phi_- + \phi_+)p_{s_0} + \phi_+ p_{s_0+1}\right], \tag{7}$$

$$\dot{p}_0 = D[p_1 - 2p_0 + p_{N-1}], \tag{8}$$

$$\dot{p}_{N-1} = D[p_{N-2} - 2p_{N-1} + p_0], \tag{9}$$

where Eqs. (8) and (9) are periodic boundary conditions, e.g. when the adatom hops off of the right side of the system ($j = N - 1$) it reappears on the left side of the lattice (at $j = 0$). Eqs. (5), (8), and (9) encode Rule 2, while Eqs. (6) and (7) encode Rules 3 and 4, respectively. Rule 1 is expressed by each of Eqs. (5)–(9), since $\dot{p}_j(t)$ only depends on $p_j(t)$ and $p_{j \pm 1}(t)$.

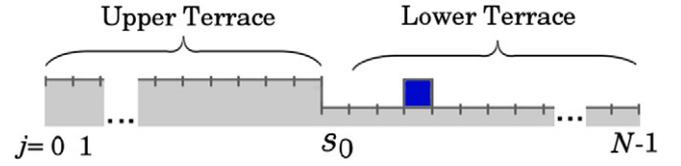

Fig. 3. The 1-p KRSOS model. A single atom [blue (dark)] is allowed to hop on the surface, whose lattice sites are indexed by $j$, where $0 \leq j \leq N - 1$. When the atom is at site $s_0$ it is a part of the step; otherwise, it is an adatom. For interpretation of the references to color in this figure legend, the reader is referred to the web version of this article.

The parameters $D$, $\phi_\pm$, and $k$ are often expressed as Arrhenius functions of the temperature [16,18]. Specifically,

$$D = \tau^{-1} e^{-E_h/k_B T}, \tag{10}$$

$$\phi_\pm = e^{-E_\pm/k_B T}, \tag{11}$$

$$k = e^{-E_b/k_B T}, \tag{12}$$

where $k_B T$ is the temperature in units of energy, $\tau^{-1}$ is a hopping frequency that is usually assumed to be $10^{13}$ s$^{-1}$, $E_h$ is an energy barrier to adatom hopping, $E_\pm$ is an attachment barrier from the left (−) or right (+) of the step,[6] and $E_b$ is a bond energy (i.e. the energy increase of the system when an adatom detaches from the step). Physically, Eqs. (10)–(12) arise from the idea that an adatom must overcome an energy barrier in order to move to a new lattice site. In particular, an adatom that attaches to a step may form a bond that must be broken in a subsequent detachment process. The barriers $E_\pm$ account for the idea that adatom motion can be hindered when attaching to a step from above or below.

Eqs. (5)–(9) are supplemented by initial data satisfying the condition $\sum_j p_j(0) = 1$; i.e. there is unit probability of finding the particle somewhere on the surface. Summing Eqs. (5)–(9) then implies that $\sum_j p_j(t) = 1$ for all times (probability is conserved). It is also possible to show that (i) the system satisfies ergodicity (any configuration is accessible to any other configuration in a finite number of transitions), (ii) there is a unique solution for any real initial data, and (iii) all initial data converge to the same steady state solution in the long-time limit. See Ref. [5] regarding a proof of statements (i)–(iii). It is also possible to solve Eqs. (5)–(9) *exactly* for all $t$, although we do not pursue this goal further. In the next section, we propose a procedure for averaging the master equation in a way that yields the BCF model.

## 4. Averaging the KRSOS Model

Our goal in this section is to develop a suitable procedure by which we can transform the KRSOS master equation into a form resembling Eqs. (1)–(4). We must reconcile two differences between the KRSOS model and the BCF theory: (i) the adatom and step positions are represented as discrete quantities in the KRSOS model but continuous variables in the BCF theory, and (ii) coordinates on the terrace are represented by a discrete index $j$ in the KRSOS model but a continuous variable $x$ in the BCF theory. In this section, we address the first of these differences.

---

[4] The actual value of $s_0$ is not important for our derivation. In fact, $s_0$ does not appear in the final form of the step velocity law.
[5] Later we will define the step position in the BCF theory as the expected microscopic step position.

[6] The barrier for an adatom to attach to the step from the upper terrace (in this case, $E_-$) is referred to as the Ehrlich–Schwoebel barrier; see also Refs. [18,19].

In statistical mechanics, measurable quantities are often defined as expectation values taken over an appropriate probability distribution, e.g. the Boltzmann distribution. Importantly, the expectation value of a random variable can be a continuous quantity, even though the random variable itself may only take discrete values. We employ this idea in defining the step position and adatom density of the BCF theory.

We begin by noting that $\dot{p}_j = 0$ if $p_j = k/Z$ for $j \neq s_0$ and $p_{s_0} = 1/Z$, where $Z = [1 + (N-1)k]$ is a normalization constant. In light of Eq. (12), we conclude that this solution is in fact the Boltzmann distribution, which we now denote $p_j^{eq}$; consequently, $Z$ is the partition function. Since the steady state is unique, *the system always approaches equilibrium at long times*.

These observations motivate us to define *time-dependent* expectation values $\varsigma(t)$ and $c_j(t)$ of the microscopic step position and adatom number-density at site $j$:

$$\varsigma(t) = \left[\sum_{j \neq s_0} a(s_0 - 1) p_j(t)\right] + a s_0 p_{s_0}(t), \qquad (13)$$

$$c_j(t) = p_j(t)/a, \qquad j \neq s_0, \qquad (14)$$

where $a = L/N$, and $L$ is the length of the system. As $t \to \infty$, the step position and adatom densities converge to their respective equilibrium expectation values, so that definitions (13) and (14) extend the notion of ensemble averaging to out-of-equilibrium systems. Importantly, this averaging procedure maps the discrete step positions $s_0$ and $s_0 - 1$ in the atomistic model to a continuous quantity $\varsigma(t)$ [because the $p_j(t)$ are continuous].

If we apply a time derivative to Eqs. (13) and (14) and use Eqs. (5)–(9) to simplify the resulting expressions, we find the (discrete) step velocity law,

$$\dot{\varsigma}(t) = a^2 D \phi_- \left(c_{s_0-1} - k p_{s_0}/a\right) \\ + a^2 D \phi_+ \left(c_{s_0+1} - k p_{s_0}/a\right), \qquad (15)$$

and discrete diffusion-type equation,

$$\dot{c}_j(t) = D\left[c_{j-1} - 2c_j + c_{j+1}\right], j \neq s_0, s_0 \pm 1, \qquad (16)$$

$$\dot{c}_{s_0 \pm 1} = D\left[k \phi_\pm c_{s_0} - (\phi_\pm + 1) c_{s_0 \pm 1} + c_{s_0 \pm 2}\right], \qquad (17)$$

$$\dot{c}_{s_0} = D\left[\phi_+ c_{s_0+1} - 2k(\phi_+ + \phi_-) c_{s_0} + \phi_- c_{s_0-1}\right]. \qquad (18)$$

Eq. (16) already has the form of Eq. (1) if we identify $\partial_x^2$ with the second-order difference scheme. However, Eqs. (17) and (18) do not have the same structure as Eq. (16); this fact allows us to determine boundary conditions for $c_j(t)$ corresponding to Eq. (2). Specifically, we add and subtract a new quantity $Dc_{s_0}^\pm$ to Eq. (17) to force the appearance of a second-order difference scheme plus some corrections, the latter of which we then require to vanish. This procedure may be interpreted as picking the boundary conditions for $c_1(t)$ such that the discrete diffusion-type Eq. (16) is valid all the way up to the step. Physically, then, we identify $c_{s_0}^\pm$ as the discrete analogues of $c^\pm$ appearing in Eq. (2).

Following through with this procedure yields

$$\dot{c}_{s_0 \pm 1} = D\left(c_{s_0}^\pm - 2c_{s_0 \pm 1} + c_{s_0 \pm 2}\right) \\ + D\left[(1 - \phi_\pm) c_{s_0 \pm 1} + (k \phi_\pm) p_{s_0}/a - c_{s_0}^\pm\right], \qquad (19)$$

where we treat the second line as the remainder term. Setting this equal to zero yields the discrete kinetic relations

$$\mp J_\pm = aD\left[c_{s_0 \pm 1} - c_{s_0}^\pm\right] = Da\phi_\pm \left[c_{s_0 \pm 1} - k p_{s_0}/a\right], \qquad (20)$$

where we identify $J_\pm = \pm (a^2 D)\left[c_{s_0 \pm 1} - c_{s_0}^\pm\right]/a$ as the discrete flux of adatoms on the right $(+)$ and left $(-)$ of the step. Note that $J_\pm$ corresponds to $\mathcal{J}_\pm = -\mathcal{D} \partial_x c(x,t)$, but with a first-order difference scheme instead of a partial derivative in $x$.

## 5. Maximum principle

At this point, we have all of the essential ingredients from which to derive the BCF theory: a step velocity law [Eq. (15)], a discrete diffusion equation [Eq. (16)], and the discrete kinetic relations [Eq. (20)]. It is tempting to take the continuum limit $a \to 0$, but we must first acknowledge that Eq. (20) has no (constant) term corresponding to $c^{eq}$; the only possible candidate is $k p_{s_0}/a$, which is a function of time.

This observation motivates the following idea: if $p_{s_0}$ remains approximately constant for all times, then to good approximation, the term $k p_{s_0}$ in Eq. (20) can be replaced by a constant that we identify as the discrete analogue of $c^{eq}$ in the linear kinetic relation [Eq. (2)]. To this end, we invoke a *maximum principle* (cf. Appendix A):

**Proposition.** *Let $p_j(t)$ be the solution to Eqs. (5)–(9) which is initially $p_j(0)$, and define $\hat{p}_j = p_j/k$ for $j \neq s_0$ and $\hat{p}_{s_0} = p_{s_0}$. Then the greatest value in the set $\{\hat{p}_j(t)\}_{j=0}^{N-1}$ is less than or equal to the greatest value in the set $\{\hat{p}_j(0)\}_{j=0}^{N-1}$ for every $t > 0$.*

Physically speaking, this maximum principle states that the (rescaled) $\hat{p}_j(t)$ will not spontaneously form localized regions with high probabilities of finding an adatom. For our purposes, it implies the following corollary:

**Corollary.** *If $p_j(t) \leq \mathcal{O}(k/Z)$ for $j \neq s_0$ and $p_{s_0}(t) = \mathcal{O}(1/Z)$ at $t = 0$, then these relations persist for all times.*

In essence, the corollary allows us to identify $p_{s_0}(t) = 1 - \mathcal{O}(kN)$ in Eq. (20) for all times, provided that $p_j(0) = \mathcal{O}(k/Z)$ for $j \neq s_0$ and $kN \ll 1$ [recall $Z = 1 + (N-1)k$]. Physically, the corollary states that if the system starts sufficiently close to equilibrium (i.e. the Boltzmann distribution), it will remain so for all times. The additional constraint $kN \ll 1$ implies a low total number of adatoms on the surface for all times. *We henceforth refer to the hypotheses about $p_j(t)$ in the above corollary as near-equilibrium conditions.*[7]

## 6. Continuum limit and the low-density regime

In order to take the continuum limit of Eqs. (15), (16), and (20), we first identify the macroscopic parameters

$$\mathcal{D} = a^2 D, \qquad (21)$$

$$c^{eq} = k/a, \qquad (22)$$

$$\kappa_\pm = aD\phi_\pm. \qquad (23)$$

---

[7] In the 1-p model, an edge atom is not available to detach from the step if an adatom is already on the surface; the term $p_{s_0}$ in Eq. (20) accounts for this fact. Such a scenario differs from real crystals, where atoms are always able to detach from a step. Nonetheless, for a corresponding model that allows for multiple detachments (e.g. the m-p model in Ref. [5]), similar near-equilibrium conditions are needed to bound corrections to the BCF theory.

We also enforce near-equilibrium conditions on the $p_j(0)$ and require $kN \ll 1$. Next, we assume that finite differences can be approximated in terms of a continuous density $c(x)$ via

$$\frac{c_{j+1}(t) - c_j(t)}{a} = \partial_x c(x,t)\big|_{x=ja} + \mathcal{O}(a), \qquad (24)$$

$$\frac{c_{j+1}(t) - 2c_j(t) + c_{j-1}(t)}{a^2} = \partial_{xx} c(x,t)\big|_{x=ja} + \mathcal{O}(a). \qquad (25)$$

Neglecting corrections that are $\mathcal{O}(a)$ or $\mathcal{O}[(kN)^2]$, we find that Eq. (15) reduces to the continuum step velocity law [Eq. (4)], Eq. (16) reduces to a diffusion equation [Eq. (1)], and Eq. (20) reduces to the linear kinetic relation at a step edge [Eq. (2)].

Because of the identity $\kappa_\pm = aD\phi_\pm$, the boundary conditions at the step edge depend strongly on the behavior of $\phi_\pm$ as $N \to \infty$. In the case that $\phi_\pm = \mathcal{O}(1)$ as $N \to \infty$, $\kappa_\pm \to \infty$, which forces the Dirichlet boundary condition $c^\pm = c^{eq}$; physically, this boundary condition corresponds to *diffusion-limited kinetics*, in which adatom hopping (as opposed to attachment/detachment) is the rate limiting process for the system to reach equilibrium.[8] If, however, $\phi_\pm = \mathcal{O}(1/N)$ as $N \to \infty$, $\kappa_\pm$ remains finite, and Eq. (2) emerges in its full form.

A key benefit of this limiting procedure is the identification of the parameters entering the BCF theory ($\mathcal{D}$, $\kappa_\pm$, and $c^{eq}$) with the parameters of the microscopic model ($D$, $k$, $\phi_\pm$, and $a$). It is possible to test the correctness of these relations via kinetic Monte Carlo simulations. The main idea of such simulations is to follow many ($10^6$ or more) elements of the statistical ensemble describing the system and then compute ensemble averages with respect to those elements; assuming that one samples enough elements of the ensemble, the simulated averages should approximate the true ensemble averages.

In Fig. 4 we compare our predictions of the parameters in the linear kinetic relation (solid line) versus kinetic Monte Carlo (KMC) simulations (points) whose input parameters are the same as the atomistic KRSOS model. *Importantly, our BCF-type model has no free parameters, since* Eqs. (21)–(23) *are determined entirely by the microscopic parameters of the KMC simulations.* The figures show that when $c \approx c^{eq}$, the linear kinetic relation $\mathcal{J}_\pm = \mp \kappa_\pm (c^\pm - c^{eq}) = \mp(aD\phi_\pm)[c^\pm - k/a]$ describes the atomistic behavior of the system remarkably well.

## 7. Discussion

### 7.1. Derivation in the context of real materials

The derivation of our BCF-type model depends on several assumptions about the scaling of the microscopic parameters: (i) $kN \ll 1$ (ii) $D = \mathcal{O}(N^2)$ s$^{-1}$, and (iii) $N \gg 1$. In this section, we consider the validity of these assumptions in the context of real materials.

The hopping rate $D$ is usually defined as the Arrhenius function $D = \tau^{-1} e^{-E_h/k_B T}$, where $\tau^{-1} = 10^{13}$ s$^{-1}$ is an attempt frequency and $E_h$ is an activation energy that is extracted from measurements [16,22]. To make contact with our model, we neglect such issues as exchange (rather than direct hopping) and multisite jumps [22]. Some typical values for $E_h$ are 0.04 eV for Al(111), 0.43 ± 0.02 eV for Cu(100), 0.94 ± 0.03 eV for W(110), and 0.97 ± 0.07 eV for Si(111) [16,22]. At temperatures between 300 K and 1000 K, we estimate that $10^{12}$ s$^{-1} \geq D \geq 10^6$ s$^{-1}$, depending on the material. As an example, we consider Ni(110), for which $E_h = 0.41$ eV [16,23][9]; taking $T \approx 500$ K (or $k_B T \approx 1/24$ eV), we estimate that $D = 10^8$ s$^{-1}$. For a terrace with $N = 1000$ lattice sites and $L = 0.1$ μm (i.e. atomic length $a \sim 0.1$ nm), we find $\mathcal{D} = D/a^2 = 1$ μm$^{-2}$ s$^{-1}$.

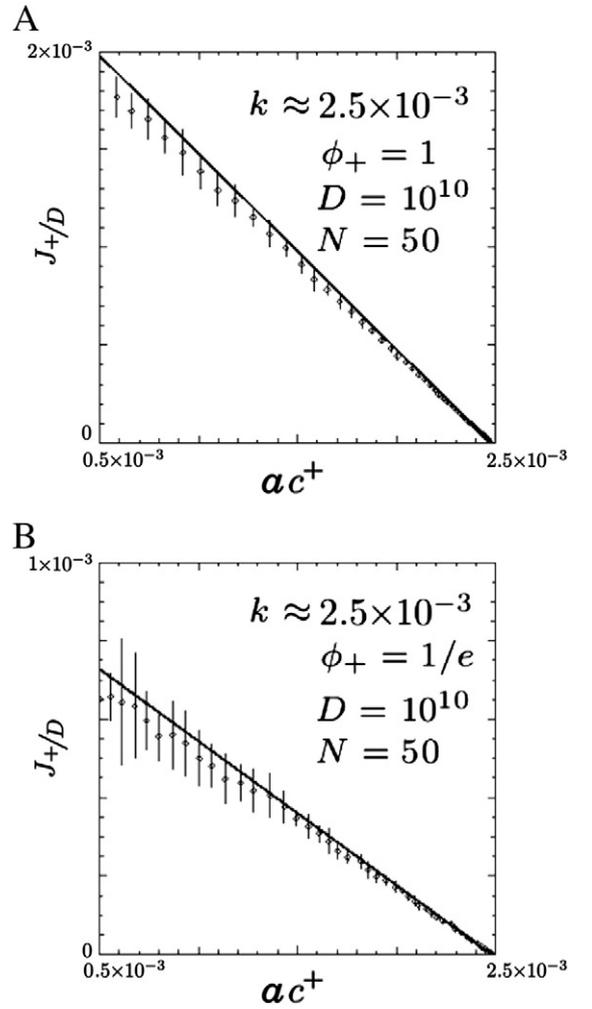

**Fig. 4.** Comparison of linear kinetic relation (2) (solid line) and kinetic Monte Carlo simulations (data points with error bars). (Top) Simulations with $\phi_+ = 1$, i.e. $E_+ = 0$. (Bottom) Simulations with $\phi_+ = 1/e$, i.e. $E_+/k_B T = 1$. We take the mean of 10 ensemble averages, with each ensemble consisting of: (a) $10^7$ simulations, and (b) $10^6$ simulations. The $3\sigma$ values are indicated by vertical lines (error bars) centered at the mean flux values. In both plots, we take $k \approx 0.0025$, $D = 10^{10}$ s$^{-1}$, and $N = 50$ (so $a = L/50$; note that $ac^{eq} = k$. The slopes of the solid lines are (a) $J_+/[D(ac^+ - k)] = -1$ and (b) $J_+/[D(ac^+ - k)] = -1/e$, in agreement with our BCF-type model. See Section 6 for a discussion.

Experiments can also estimate the energy $E_b$ [cf. Eq. (12)]. Typical values range from approximately 0.3 eV for Ni(110) [23] up to 1 eV or 2 eV for Si(111) [24–26]. The use of the value $E_b = 0.3$ eV for Ni(110) [cf. Eq. (12)] yields $k \approx 10^{-4}$ at 500 K. By combining this result with the assumption that $N = 1000$ (corresponding to $L$ that is a few hundred nanometers), we find that $Nk \approx 10^{-1}$, which suggests that the low-density approximation is reasonable for this system at 500 K. In addition to these formal estimates, both experimental and numerical results have verified that Ni(110) is in a low-density regime at this temperature; see Ref. [23]. In this work, significant adatom detachment on Ni(110) only began when the temperature was raised above 650 K; at

---

[8] Derivations of BCF-type equations from phase-field models (e.g. as in Ref. [20]) suggest that $\kappa_\pm = aD(\phi_\pm - 1)$, which forces the Dirichlet boundary condition when $\phi_\pm \to 1$. In Eq. (23), there is no need for the $-1$ term since the scaling of $\phi_\pm$ relative to $N$ is sufficient to determine the step kinetics.

[9] We caution that Ni(110) and other (110) fcc metals, as well as (211) bcc metals, are anisotropic, with lower barriers in channel than cross channel when exchange processes are insignificant [21]. Such processes are beyond the scope of our simple BCF-type theory. With the quoted energy values, we seek to convey the order of magnitude of the key parameters [15,21].

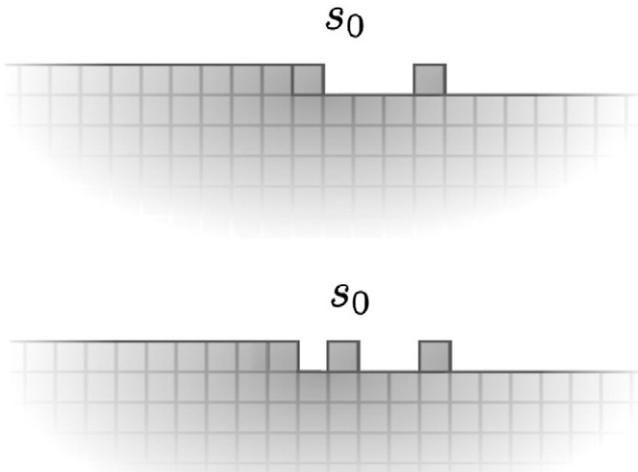

**Fig. 5.** A convection-type process that contributes corrections to the BCF theory. Here the adatom moved relative to the step because another atom detached from the step.

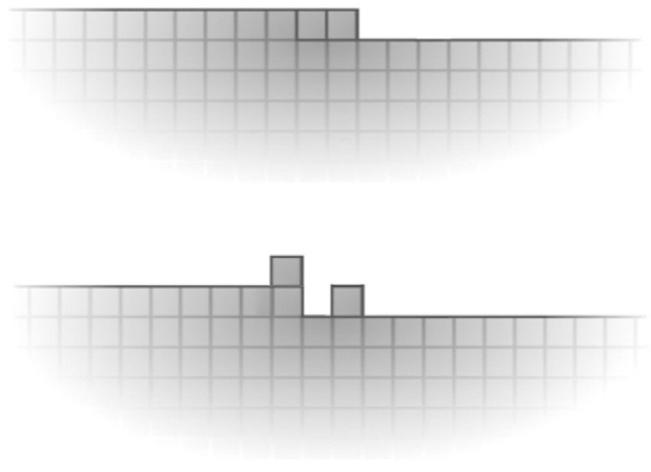

**Fig. 6.** Another process that yields corrections to the BCF theory: a transition between states in which the step changes its position by more than one lattice site.

900 K, simulations show that roughly 1.5% of the lattice sites are occupied by adatoms (see also Ref. [17]).[10]

Experimental estimates of $E_\pm$ are also available [cf. Eq. (11)]. Often (but not always) the Ehrlich–Schwoebel barrier [19,20] $E_-$ is larger than the attachment barrier $E_+$. See, e.g. Table 6 in Ref. [16] for a detailed list of attachment/detachment barriers. For Ni(110), one finds $E_- = 0.9$ eV and $E_+ \approx 0$ eV, which implies $\phi_- \ll 1/N$ and $\phi_+ = 1$ at 500 K. In a BCF model for this system, we therefore expect that $\kappa_- \approx 0$ and $\kappa_+ = \mathcal{O}(N)$, corresponding to $\mathcal{J}_- = 0$ and $c_+ = c^{eq}$. Therefore, for this system, our analysis predicts different boundary conditions on the two sides of the step edge.

### 7.2. Corrections due to multiple-particle states

The KRSOS master equation that we invoke in Section 3 only allows a single adatom to ever be on the surface. We comment here on corrections that can arise between multi-adatom states. Note that generally, the particular form of corrections will depend on precise rules of the multi-particle KRSOS model; here we discuss only a few types. However, we always expect that an m-adatom state should be $\mathcal{O}[(kN)^m]$ (provided the system is near-equilibrium), since $mE_b$ is the energy cost to create $m$ adatoms. Thus, corrections from multi-particle states should always be small; see Ref. [5].

In Fig. 5 transitions between the two states illustrated cause the adatom furthest from the step to move *relative to the step*. This is an example of a convection-type process. If the definition of the adatom density in a multi-particle model is taken relative to the step position, then such convective effects will appear in the diffusion equation for adatoms. Importantly, this process will not contribute a simple convection term $\dot{\varsigma}\partial_x c(x,t)$ to the diffusion Eq. (1) because the probabilities that *both* an adatom is at some site *and* another atom detaches from the step are not independent; see Ref. [5].

In Fig. 6, transitions between the two states illustrated cause the step to move forward or backward by two lattice sites, as opposed to one. Any such process in which a single attachment (or detachment) event causes the step to move by more than one lattice site will introduce corrections to the step velocity law. Formally, this can be understood by examining Eq. (4): a single factor of $a$, as opposed to $2a, 3a,...,$ multiplies the net flux to the step. In the BCF theory, attachment of adatoms to a step causes it to move by a fixed distance.

### 7.3. Limitations

Our KRSOS model has limitations because we only consider a single step in 1D. In this setting, it is not possible to derive step interactions. In many formulations of the BCF theory, such interactions introduce an additional energy into the step chemical potential, so that the energy cost of adatom detachment depends on the widths of the terraces adjacent to the step [28,29,16]. We speculate that in an appropriate multi-step KRSOS model, this energy penalty should appear as an additional, configuration-dependent contribution to $E_b$.

Because our KRSOS model is only 1D, we cannot account for the effects of anisotropy in the crystal lattice. Such effects could be important in systems such as Si(001), where diffusion rates depend on both direction and position [24,25]. We speculate that an appropriate atomistic model incorporating these features would lead to a BCF model with an anisotropic and (potentially) position dependent diffusion coefficient.

Our analysis is also unable to determine the role that kinks (i.e. bends in the step) play in the derivation of BCF-type models. In 2D kinetic solid-on-solid models, it is known that kinks, which alter the microscopic step profile, play an important role in determining the rates of adatom attachment/detachment processes [11,14]. Moreover, in 2D BCF-type models, the step chemical potential (i.e. the energy cost to remove an adatom from a step), and consequently the linear kinetic relations are typically assumed to depend on the local step curvature [3,16,28].

## 8. Conclusions

In this paper, we showed how a 1D time-dependent, single-step version of the BCF model can be derived from an atomistic, stochastic lattice-gas model of the surface. We used an averaging procedure to connect the atomistic system configurations to the notions of a continuous adatom density and step position; we then showed that the BCF theory (plus corrections) describes the time evolution of these averages. Via a maximum principle, we also showed that corrections are negligible when the system is sufficiently close to equilibrium. Our use of an averaging procedure and maximum principle reveals the sense in which the BCF theory can be viewed as a model of mesoscale phenomena in non-equilibrium statistical mechanics.

---

[10] There has been recent interest in systems, such as metal nitrides, with much higher characteristic energies. While such systems introduce 2-species issues, density-functional theory (DFT) calculations [26] show that TiN and ScN have cohesive energies higher than 7 eV. For such systems BCF theory might be valid at temperatures as high as 1000 K.

### Note added in proof

We recently became aware that a derivation of a relation similar to Eq. (20) is given by Pimpinelli and Villain [30] by use of a detailed balance argument for continuum terrace diffusion; see their Eqs. (6.17). We believe that our viewpoint and method are different, relying entirely on an atomistic scheme.

### Acknowledgments

PNP was supported by the National Institute of Standards and Technology American Recovery and Reinvestment Act Measurement Science and Engineering Fellowship Program Award No. 70NANB10H026 through the University of Maryland, with ancillary support from the NSF-MRSEC under Grant No. DMR 05-20471. This author's research was also supported by NSF-DMS 08-47587 at the University of Maryland and the Institute for Mathematics and its Applications at the University of Minnesota. TLE was supported in part by the NSF-MRSEC Grant DMR 05-20741, and by NSF-CHE 07-50334 and 13-05892. DM was supported by NSF-DMS 08-47587 at the University of Maryland. We acknowledge helpful interactions with A. BH. Hammouda, M. K. Hawkins, J. A. Liddle, O. Pierre-Louis, T. Schulze, M. Stiles, and J. Weare.

### Appendix A. Proof of the maximum principle

**Proposition.** *Let $p_j(t)$ be the solution to* Eqs. (5)–(9) *which is initially $p_j(0)$, and define $\hat{p}_j = p_j/k$ for $j \neq s_0$ and $\hat{p}_{s_0} = p_{s_0}$. Let* max $\{\cdot\}$ *be the greatest value in the set* $\{\cdot\}$. *Then $\hat{p}_j$ satisfies the maximum principle that* $\max_j\{\hat{p}_j(t)\} \leq \max_j\{\hat{p}_j(0)\}$ *for all $t>0$.*

**Proof.** We proceed by *reductio ad absurdum*. Writing Eqs. (5)–(9) in terms of $\hat{p}_j$ yields

$$k\dot{\hat{p}}_j = Dk\left[\hat{p}_{j+1} - 2\hat{p}_j + \hat{p}_{j-1}\right], \qquad j \neq s_0, s_0 \pm 1,$$
$$k\dot{\hat{p}}_{s_0 \pm 1} = Dk\left[\phi_\pm \hat{p}_{s_0} - (1+\phi_\pm)\hat{p}_{s_0 \pm 1} + \hat{p}_{s_0 \pm 2}\right], \qquad (A.1)$$
$$\dot{\hat{p}}_{s_0} = Dk\left[\phi_- \hat{p}_{s_0 - 1} - (\phi_- + \phi_+)\hat{p}_{s_0} + \phi_+ \hat{p}_{s_0 + 1}\right].$$

Assume that at some time $t$ there is an $l$ such that $\hat{p}_l(t) \geq 0$ and $\hat{p}_l(t) \geq \hat{p}_j(t)$ for all $j \neq l$. But by virtue of Eq. (A.1), we obtain

$$\hat{p}_l(t) \geq \frac{\theta_1 \hat{p}_{l-1}(t) + \theta_2 \hat{p}_{l+1}(t)}{\theta_1 + \theta_2},$$

where $\theta_{1,2}$ are 1 or $\phi_\pm$, depending on the value of $l$. By assumption, it is impossible that $\hat{p}_{l\pm 1}(t) > \hat{p}_l(t)$, so that either $\hat{p}_l$ is not a maximum or $\hat{p}_j$ is constant for all $j$.

### References

[1] C. Misbah, O. Pierre-Louis, Y. Saito, Rev. Mod. Phys. 82 (2010) 981.
[2] U.S. Department of Energy Report, , Argonne National Laboratory, Argonne, IL, 2012. (Available online at http://science.energy.gov/bes/news-and-resources/reports/basic-research-needs/ ).
[3] W.K. Burton, N. Cabrera, F.C. Frank, Philos. Trans. R. Soc. Lond. Ser. A 243 (1951) 299.
[4] R. Ghez, S.S. Iyer, IBM J. Res. Dev. 32 (1988) 804.
[5] P.N. Patrone, D. Margetis, Multiscale Model. Simul. 12 (1) (2014) http://epubs.siam.org/doi/abs/10.1137/13091587X .
[6] B. Krishnamachari, J. McLean, B. Cooper, J. Sethna, Phys. Rev. B 54 (1996) 8899.
[7] J.D. Weeks, G.H. Gilmer, in: I. Prigogine, S.A. Rice (Eds.), Advances in Chemical Physics, vol. 40, John Wiley, New York, 1979, p. 157.
[8] M. Ozdemir, A. Zangwill, Phys. Rev. B 45 (1990) 3718.
[9] W.A. Strauss, Partial Differential Equations: An Introduction, Wiley, Hoboken, 2008. 41.
[10] J.G. McLean, B. Krishnamachari, D.R. Peale, E. Chason, J.P. Sethna, B.H. Cooper, Phys. Rev. B 55 (1997) 1811.
[11] D. Ackerman, J. Evans, Multiscale Model. Simul. 9 (2011) 59.
[12] M. Saum, T.P. Schulze, Discret. Cont. Dyn. B 11 (2009) 443.
[13] T.P. Schulze, J. Cryst. Growth 295 (2006) 188.
[14] A. Zangwill, C.N. Luse, D.D. Vvedensky, M.R. Wilby, Surf. Sci. 274 (1992) L529.
[15] T. Zhao, J.D. Weeks, D. Kandel, Phys. Rev. B 71 (2005) 155326.
[16] H.-C. Jeong, E.D. Williams, Surf. Sci. Rep. 34 (1999) 171.
[17] A.BH. Hamouda, A. Pimpinelli, T.L. Einstein, Surf. Sci. 602 (2008) 3569.
[18] A.F. Voter, in: K.E. Sickafus, E.A. Kotomin, B.P. Uberuaga (Eds.), Radiation Effects in Solids, ATO Sci. Series, vol. 235, Springer, Netherlands, 2007, p. 1.
[19] G. Ehrlich, F.G. Hudda, J. Chem. Phys. 44 (1966) 1039.
[20] R.L. Schwoebel, E.J. Shipsey, J. Appl. Phys. 37 (1966) 3682.
[21] O. Pierre-Louis, Phys. Rev. E. 68 (2003) 021604.
[22] G. Antczak, G. Ehrlich, Surf. Sci. Rep. 62 (2007) 39;
G. Antczak, G. Ehrlich, Surface Diffusion: Metals, Metal Atoms, and Clusters, Cambridge Univ. Press, Cambridge, 2010.
[23] W. Silvestri, A.P. Graham, J.P. Toennies, Phys. Rev. Lett. 81 (1998) 1034.
[24] C. Roland, G.H. Gilmer, Phys. Rev. B 46 (1992) 13428.
[25] C. Roland, G.H. Gilmer, Phys. Rev. B 46 (1992) 13437.
[26] S. Kodiyalam, K.E. Khor, S. Das Sarma, Phys. Rev. B 53 (1996) 9913.
[27] Z.T.Y. Liu, X. Zhou, S.V. Khare, D. Gall, J. Phys. Condens. Matter 26 (2014) 025404.
[28] O. Pierre-Louis, C. Misbah, Phys. Rev. B 58 (1998) 2259.
[29] P.N. Patrone, T.L. Einstein, D. Margetis, Phys. Rev. E. 82 (2010) 061601.
[30] A. Pimpinelli, J. Villain, Physics of Crystal Growth, Cambridge University Press, Cambridge, UK, 1998. pp. 96–99.